\documentclass[sigconf]{acmart}
\usepackage[utf8]{inputenc} % allow utf-8 input
\usepackage[T1]{fontenc}    % use 8-bit T1 fonts
\usepackage{hyperref}       % hyperlinks
\usepackage{url}            % simple URL typesetting
\usepackage{amsfonts}       % blackboard math symbols
\usepackage{nicefrac}       % compact symbols for 1/2, etc.
\usepackage{microtype}      % microtypography
\usepackage{xcolor}         % colors
\usepackage{amsthm}
\usepackage{subfigure}
\usepackage{cleveref}
\usepackage{comment}
\usepackage{multirow}
\usepackage{enumitem}
\usepackage{wrapfig}
\usepackage{balance}
\usepackage{bbding}
\usepackage{adjustbox}
\usepackage{bbding}
\newtheorem{definition}{Definition}
\definecolor{Instruction}{RGB}{237, 84, 102}
\definecolor{Example}{RGB}{81, 140, 192}
\definecolor{Question}{RGB}{0, 0, 0}
\definecolor{Answer}{RGB}{140, 192, 81}

\AtBeginDocument{%
  \providecommand\BibTeX{{%
    \normalfont B\kern-0.5em{\scshape i\kern-0.25em b}\kern-0.8em\TeX}}}

\copyrightyear{2023}
\acmYear{2023}
\setcopyright{rightsretained}
\acmConference[RecSys '23]{Seventeenth ACM Conference on Recommender Systems}{September 18--22, 2023}{Singapore, Singapore}
\acmBooktitle{Seventeenth ACM Conference on Recommender Systems (RecSys '23), September 18--22, 2023, Singapore, Singapore}\acmDOI{10.1145/3604915.3610646}
\acmISBN{979-8-4007-0241-9/23/09}

\begin{document}

\title{Uncovering ChatGPT's Capabilities in Recommender Systems}
% \setCJKfamilyfont{font01}{AR PL KaitiM GB}
% \affiliation{%

\author{Sunhao Dai}
\authornote{The first three authors contributed equally.}
\affiliation{
\institution{Gaoling School of Artificial Intelligence, Renmin University of China}
\city{Beijing}\country{China}
}
\email{sunhaodai@ruc.edu.cn}

\author{Ninglu Shao}
\authornotemark[1]
\affiliation{%
\institution{Gaoling School of Artificial Intelligence, Renmin University of China}
\city{Beijing}\country{China}
}
\email{ninglu_shao@ruc.edu.cn}

\author{Haiyuan Zhao}
\authornotemark[1]
\affiliation{
\institution{School of Information, Renmin University of China}
\city{Beijing}\country{China}
}
\email{haiyuanzhao@ruc.edu.cn}

\author{Weijie Yu}
\affiliation{
\institution{School of Information Technology and Management, University of International Business and Economics}
\city{Beijing}\country{China}
}
\email{yuweijie23@gmail.com}

\author{Zihua Si}
\affiliation{%
\institution{Gaoling School of Artificial Intelligence, Renmin University of China}
\city{Beijing}\country{China}
}
\email{zihua_si@ruc.edu.cn}

\author{Chen Xu}
\affiliation{%
\institution{Gaoling School of Artificial Intelligence, Renmin University of China}
\city{Beijing}\country{China}
}
\email{xc_chen@ruc.edu.cn}

\author{Zhongxiang Sun}
\affiliation{%
\institution{Gaoling School of Artificial Intelligence, Renmin University of China}
\city{Beijing}\country{China}
}
\email{sunzhongxiang@ruc.edu.cn}

\author{Xiao Zhang}
\affiliation{%
\institution{Gaoling School of Artificial Intelligence, Renmin University of China}
\city{Beijing}\country{China}
}
\email{zhangx89@ruc.edu.cn}

\author{Jun Xu}
\authornote{Jun Xu is the corresponding author. Work partially done at Engineering Research Center
of Next-Generation Intelligent Search and Recommendation, Ministry of Education.
}
\affiliation{%
  \institution{Gaoling School of Artificial Intelligence, Renmin University of China}
   \city{Beijing}\country{China}
  }
\email{junxu@ruc.edu.cn}

\renewcommand{\shortauthors}{Sunhao Dai et al.}


\begin{abstract}
The debut of ChatGPT has recently attracted significant attention from the natural language processing (NLP) community and beyond. Existing studies have demonstrated that ChatGPT shows significant improvement in a range of downstream NLP tasks, but the capabilities and limitations of ChatGPT in terms of recommendations remain unclear.   
In this study, we aim to enhance ChatGPT's recommendation capabilities by aligning it with traditional information retrieval (IR) ranking capabilities, including point-wise, pair-wise, and list-wise ranking. To achieve this goal, we re-formulate the aforementioned three recommendation policies into prompt formats tailored specifically to the domain at hand.  Through extensive experiments on four datasets from different domains, we analyze the distinctions among the three recommendation policies. Our findings indicate that ChatGPT achieves an optimal balance between cost and performance when equipped with list-wise ranking. This research sheds light on a promising direction for aligning ChatGPT with recommendation tasks. To facilitate further explorations in this area, the full code and detailed original results are open-sourced at \url{https://github.com/rainym00d/LLM4RS}.
\end{abstract}


%\keywords{learning to rank, examination bias, trust bias, joint learning}


\maketitle

\section{Introduction}\label{Introduction}

\begin{figure*}[t]  
    \centering    
    \includegraphics[width=1\linewidth]{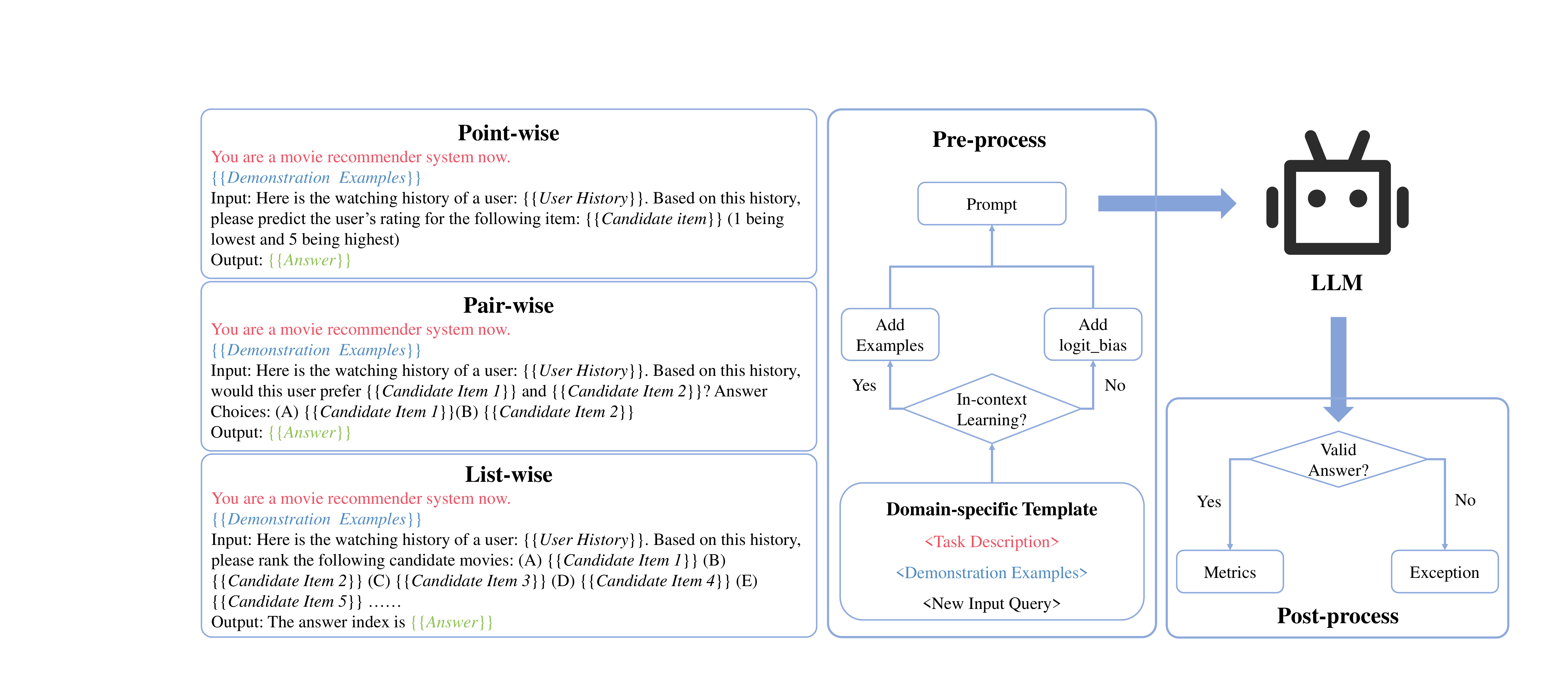}
    \caption{The overall evaluation framework of LLMs for recommendation.
    The left part demonstrates examples of how prompts are constructed to elicit each of the three ranking capabilities.
    The right part outlines the process of employing LLMs to perform different ranking tasks and conduct evaluations.}
    \label{fig:model}  
\end{figure*}

Large language models (LLMs), such as ChatGPT developed by OpenAI~\cite{ChatGPT}, have recently gained significant attention from the natural language processing (NLP) community and beyond. These LLMs possess a versatile nature and extensive world knowledge, allowing them to be applied not only in various NLP tasks~\cite{brown2020language, kojima2022large, chowdhery2022palm}, but also in domains such as education~\cite{m2022exploring, guo2023close, Nunes2023EvaluatingGA}, medicine~\cite{antaki2023evaluating, rao2023evaluating, benoit2023chatgpt}, search~\cite{Sun2023IsCG, ma2023zero,qin2023webcpm} and law~\cite{choi2023chatgpt, cui2023chatlaw, sun2023short}. 

Meanwhile, previous research has indicated that off-the-shelf pre-trained language models (LMs) can be directly used as recommenders by adapting recommendation tasks into multi-token cloze tasks using prompts~\cite{penha2020does, sileo2022zero, zhang2021language}. Hence, a natural research question arises regarding how to effectively align LLMs with recommendation capabilities and whether LLMs can work as few-shot or zero-shot recommender systems.

The primary objective of recommender systems is to alleviate information overload by providing personalized top-$K$ item ranking lists for users~\cite{resnick1997recommender}. In information retrieval (IR), previous studies have commonly utilized three approaches to construct these ranking lists: point-wise, pair-wise, and list-wise~\cite{liu2009learning, xu2018deep}. Consequently, in this paper, our specific focus is on probing the recommendation capabilities of LLMs by aligning them with these three ranking perspectives.
Detailed formulation of these three ranking perspectives could be seen in Section~\ref{sec:methods}.

To investigate the potential of LLMs in recommendation tasks from these three ranking perspectives, we begin by reformulating the three capabilities into prompts that are tailored to the specific domain and serve as input for LLMs. We then conduct an empirical analysis of  ChatGPT and other popular LLMs from OpenAI on four widely-used recommendation benchmarks from different knowledge-rich domains. To the best of our knowledge, this is the first empirical study to probe the capabilities of ChatGPT in recommender systems from different ranking perspectives.

\textbf{Major Findings.}  In summary, we have the following major findings after empirical experiments:
\begin{itemize}  [leftmargin=0.5cm]
    \item ChatGPT shows consistent advantages in all three ranking capabilities compared with other LLMs.
    \item ChatGPT is good at list-wise and pair-wise ranking while less good at point-wise ranking. 
    \item ChatGPT can outperform traditional recommendation models with limited training data.
    \item Considering the improvements with cost, we recommend list-wise ranking for LLM-based recommenders in practice. 
    \item ChatGPT exhibits potential in explainable recommendations and a good understanding of item similarity.
\end{itemize}

We hope that this preliminary evaluation of ChatGPT in recommendation can provide new perspectives on both assessing the capabilities of LLMs and utilizing LLMs, such as ChatGPT, to enhance recommender systems.

\section{Background}

\subsection{Large Language Models}

Pioneering studies~\cite{radfordlanguage,brown2020language} demonstrated that LLMs can perform a diverse range of tasks without requiring gradient updates, solely based on textual instructions or a few examples. This has drawn significant attention towards improving the capabilities of LLMs. 
Previous studies~\cite{kaplan2020scaling} have investigated the performance limits of pre-trained language models (PLMs) by training larger models, as they have noted that augmenting the model or data size typically enhances the model's ability on downstream tasks, 
such as Megatron-turing NLG~\cite{smith2022using} with 530B parameters, Gopher~\cite{rae2021scaling} with 280B parameters,  Ernie 3.0 Titan ~\cite{wang2021ernie} with 260B parameters, BLOOM~\cite{scao2022bloom} with 175B parameters, and PaLM~\cite{chowdhery2022palm} with 540B parameters. 
These LLMs have exhibited exceptional performance on challenging tasks, showcasing new abilities that were not apparent in smaller pre-trained language models.  For a more comprehensive overview of LLMs, we would recommend referring to~\cite{zhao2023survey}.

\subsection{Language Models for Recommendation}
The remarkable success of pre-trained LMs in NLP community has motivated researchers in recommender systems to explore their potential in recommendation tasks. Existing works can be categorized into two types: (i) utilizing LMs training strategies to reformulate and model recommendation tasks, such as BERT4Rec (\textit{masked language modeling}) \cite{sun2019bert4rec},  UnisRec (\textit{pre-train and finetune})\cite{hou2022towards}, P5 (\textit{pre-train and prompting}) \cite{geng2022recommendation} and (ii) using LMs to obtain better representations of users and items as extra features based on textual information \cite{wu2021empowering}. More recently, some researchers have explored leveraging off-the-shelf pre-trained LMs as recommender systems by reformulating the recommendation tasks with prompts as multi-token cloze tasks \cite{zhang2021language,sileo2022zero,penha2020does}. 
In this paper, we aim to conduct a preliminary evaluation of ChatGPT's potential and limitations in recommender systems.

\section{Probing ChatGPT for Recommendation Capabilities} \label{sec:methods}
In this section, we leverage prompts to adapt point-wise, pair-wise, and list-wise ranking tasks, enabling off-the-shelf LLMs to effectively tackle these tasks.

\begin{table*}[]

  \caption{Overall performance of different models on four  datasets from different domains. Bold indicates the best result for each row and `\_' indicates the best result for each LLM.
  `random' denotes recommendation based on a random policy.
  `pop' denotes recommendation based on items' popularity judged by the number of interactions.
  }
    \label{tab:main_results}
    \resizebox{1\textwidth}{!}{
  \begin{tabular}{c|c|c|c|ccc|ccc|ccc}
  \toprule
  \multirow{2}{*}{Domain} & \multirow{2}{*}{Metric} & \multirow{2}{*}{random} & \multirow{2}{*}{pop} & \multicolumn{3}{c}{text-davinci-002} & \multicolumn{3}{|c|}{text-davinci-003} & \multicolumn{3}{c}{gpt-3.5-turbo (ChatGPT)} \\
  & & & & point-wise & pair-wise & list-wise & point-wise & pair-wise & list-wise & point-wise & pair-wise & list-wise \\
  \midrule
  \multirow{3}{*}{Movie} & Compliance Rate & - & - & 100.00\% & 100.00\% & 100.00\% & 100.00\% & 100.00\% & 100.00\% & 100.00\% & 99.98\% & 100.00\% \\
  & NDCG@3 & 0.4262 & 0.4761 & 0.5416 & \underline{ 0.5728} & 0.4990 & 0.4618 & 0.5441 & \underline{ 0.5564} & \underline{ \textbf{0.5912}} & 0.5827 & 0.5785 \\
  & MRR@3 & 0.3667 & 0.4103 & 0.4824 & \underline{ 0.5071} & 0.4363 & 0.3998 & 0.4763 & \underline{ 0.4950} & \underline{ \textbf{0.5260}} & 0.5162 & 0.5167 \\
  \midrule
  \multirow{3}{*}{Book} & Compliance Rate & - & - & 99.96\% & 100.00\% & 100.00\% & 100.00\% & 100.00\% & 100.00\% & 100.00\% & 99.98\% & 99.80\% \\
  & NDCG@3 & 0.4262 & 0.4999 & 0.4889 & \underline{ 0.5298} & 0.4290 & 0.4585 & \underline{ 0.5293} & 0.4597 & 0.5075 & 0.5350 & \underline{ \textbf{0.5395}} \\
  & MRR@3 & 0.3667 & 0.4340 & 0.4247 & \underline{ 0.4646} & 0.3690 & 0.3993 & \underline{ 0.4665} & 0.4040 & 0.4495 & 0.4774 & \underline{ \textbf{0.4800}} \\
  \midrule
  \multirow{3}{*}{Music} & Compliance Rate & - & - & 100.00\% & 100.00\% & 100.00\% & 100.00\% & 100.00\% & 100.00\% & 100.00\% & 99.96\% & 99.80\% \\
  & NDCG@3 & 0.4262 & 0.4094 & 0.4623 & \underline{ 0.4681} & 0.4277 & 0.4732 & \underline{ 0.5072} & 0.4506 & 0.5201 & 0.5439 & \underline{ \textbf{0.5567}} \\
  & MRR@3 & 0.3667 & 0.3470 & 0.4030 & \underline{ 0.4082} & 0.3750 & 0.4113 & \underline{ 0.4448} & 0.4000 & 0.4605 & 0.4830 & \underline{ \textbf{0.4950}} \\
  \midrule
  \multirow{3}{*}{News} & Compliance Rate & - & - & 99.80\% & 100.00\% & 100.00\% & 100.00\% & 100.00\% & 100.00\% & 100.00\% & 100.00\% & 99.60\% \\
  & NDCG@3 & 0.4262 & \textbf{0.5444} & 0.4483 & 0.4550 & \underline{ 0.5059} & 0.4880 & \underline{ 0.4892} & 0.4742 & 0.4826 & 0.4991 & \underline{ 0.5094} \\
  & MRR@3 & 0.3667 & \textbf{0.4840} & 0.3879 & 0.3936 & \underline{ 0.4497} & 0.4271 & \underline{ 0.4294} & 0.4173 & 0.4246 & 0.4354 & \underline{ 0.4515} \\       
  \bottomrule
\end{tabular}}
\end{table*}

\begin{table*}[]
\caption{Rank of different capabilities of different LLMs-based recommendation models on four  datasets from different domains.}
  \label{tab:rank}
  \resizebox{1\textwidth}{!}{
\begin{tabular}{c|c|c|c}
\toprule
Domain & text-davinci-002      & text-davinci-003      & gpt-3.5-turbo (ChatGPT)      \\
\midrule
Movie  & pair-wise $>$ point-wise $\gg$ list-wise  & list-wise $\approx$ pair-wise $\gg$ point-wise          & point-wise $>$ pair-wise $\approx$ list-wise \\
Book   & pair-wise $\gg$ point-wise $\gg$ list-wise         & pair-wise $\gg$ list-wise $\approx$ point-wise           & list-wise $>$ pair-wise $\gg$ point-wise \\
Music  & pair-wise $>$ point-wise $\gg$ list-wise         & pair-wise $\gg$ point-wise $\gg$ list-wise          & list-wise $>$ pair-wise $\gg$ point-wise \\
News   & list-wise $\gg$ pair-wise $\approx$ point-wise         & pair-wise $\approx$ point-wise $>$ list-wise    & list-wise $>$ pair-wise $>$ point-wise  \\
\bottomrule
\end{tabular}}
\end{table*}

\subsection{Three Ranking Capabilities in Recommender Systems}
The core objective of personalized recommendation is to rank candidate items based on user preferences. To accomplish this, current learning-to-rank (LTR) methods empower different capabilities to recommender systems via corresponding loss functions, including point-wise ranking capability, pair-wise ranking capability and list-wise ranking capability~\cite{liu2009learning}. Formally, given a user $u \in  \mathcal{U}$ and $k$ candidate items $\{i_1,i_2,\cdots,i_k\}\subset \mathcal{I}$, each user-item pair's representation is encoded as $\mathbf{x}_{u,i}$. The above three capabilities can be formulated as follows:

\begin{definition}{(point-wise ranking capability)}
The recommender system learns to predict the preference score of each item $i$ for each user $u$ via a point-wise scoring function $\Phi_{\mathrm{point}}(\cdot)$:
$
s(i\mid u) = \Phi_{\mathrm{point}}(\mathbf{x}_{u,i})
$
The preference score $s$ is then used to rank the items for each user. 
The common used loss function in point-wise ranking includes mean squared error (MSE)~\cite{DBLP:conf/sigir/RendleGFS11} and binary cross entropy (BCE)~\cite{he2017neural}.  
\end{definition}

\begin{definition}{ (pair-wise ranking capability)}
The recommender system learns to compare pairs of items $i_m$ and $i_n$ for each user $u$ and predict which item is preferred via a pair-wise scoring function $\Phi_{\mathrm{pair}}(\cdot)$:
$
s(i_m\succ i_n\mid u) = \Phi_{\mathrm{pair}}(\mathbf{x}_{u,i_m},\mathbf{x}_{u,i_n})
$
The system then ranks the items based on the relative preference score $s$.
The pairwise hinge loss~\cite{Joachims2002Optimizing} or Bayesian personalized ranking loss (BPR)~\cite{Rendle2009BPR} are the typical loss functions utilized in pair-wise ranking.
\end{definition}

\begin{definition}{ (List-wise ranking capability)}
The recommender system learns to directly predict the preference score of a list of items $\{i_1,i_2,\cdots,i_k\}$ for each user $u$ via a list-wise scoring function $\Phi_{\mathrm{list}}(\cdot)$:
$
s(i_1\mid u),s(i_2\mid u),\cdots,s(i_k\mid u) = \Phi_{\mathrm{list}}(\mathbf{x}_{u,i_1},\mathbf{x}_{u,i_2},\cdots,\mathbf{x}_{u,i_k})
$
The system then sorts the items based on the predicted scores. The list-wise loss, e.g., sampled softmax loss~\cite{Youtube_rec_sys_16} is employed to optimize the recommendation model.
\end{definition}

\subsection{Reformulate and Adapt Recommendation with Prompts}\label{sec: prompts}
% unified
To obtain above capabilities of recommendation, current recommendation models employ corresponding loss functions for supervised learning. However, the supervised learning schema often fails in data sparsity scenarios (e.g., cold start problems~\cite{gope2017survey} and long-tailed items~\cite{park2008long}). In contrast, LLMs have a stronger generalization capability in these data sparsity scenarios and achieve promising performances in few-shot and even zero-shot tasks. In this empirical study, we assume that LLMs already have the above three capabilities, and all we need to do is to trigger these capabilities through prompt tuning. To this end, we adopt the recent successful practice of in-context learning~\cite{brown2020language} and instruction tuning~\cite{chung2022scaling}, and we express the aforementioned three capabilities as three tasks with domain-specific prompts.

Figure~\ref{fig:model} illustrates how we employ prompt tuning to elicit three ranking capabilities from LLMs.
As shown in Figure~\ref{fig:model} (left), our prompt consists of three components: 
(i) \textit{Task description} $I$ refers to the process of enabling the LLM to comprehend the particular domain in which it is required to perform recommendation tasks.
The task description is designed to be domain-aware, which enhances LLM's perception of pertinent knowledge. (ii) \textit{Demonstration examples} $\mathcal{D}=\{f(\mathbf{h}_1,\mathbf{c}_1,\mathbf{y}_1),\cdots,f(\mathbf{h}_N,\mathbf{c}_N,\mathbf{y}_N)\}$ (i.e., $N$-shot in-context learning), where $\mathbf{h}$ denotes the historical interacted items of a user, $\mathbf{c}$ denotes the candidate items which need to be ranked, $\mathbf{y}$ denotes the predictions given by LLMs, and $f(\cdot)$ is the function for transforming the examples into designed prompt templates.
The demonstration examples facilitate the LLM's comprehension of the current task.
% \szh{Too redundant. The demonstration examples facilitate the LLM's comprehension of the current task.}
(iii) \textit{New input query} of a given user $f(\mathbf{h}^{\prime},\mathbf{c}^{\prime} \mid u)$, which needs to be answered by LLMs. 
For three ranking tasks, the corresponding candidate items $\mathbf{c}$ are constructed as follows:

\[
\begin{split}
    \mathbf{c} = \begin{cases}
    \{i\} & \text{for point-wise ranking capability,} \\
    \{i_m,i_n\} & \text{for pair-wise ranking capability,} \\
    \{i_1,i_2,\cdots,i_k\} & \text{for list-wise ranking capability.} 
    \end{cases}
\end{split}
\]

As depicted in Figure~\ref{fig:model} (right), LLMs will utilize the different ranking capabilities elicited through different prompts to make predictions  $\hat{y}^{\prime}$:
\[
\begin{split}
    \hat{y}^{\prime}_i &= LLM_{\mathrm{point}}\left(I,\mathcal{D},f(\mathbf{h}^{\prime},\mathbf{c}^{\prime}\mid u)\right), \\
    \hat{y}^{\prime}_{i_m\succ i_n} &= LLM_{\mathrm{pair}}\left(I,\mathcal{D},f(\mathbf{h}^{\prime},\mathbf{c}^{\prime}\mid u)\right), \\
    \hat{y}^{\prime}_{i_1},\hat{y}^{\prime}_{i_2},\cdots,\hat{y}^{\prime}_{i_k} &= LLM_{\mathrm{list}}\left(I,\mathcal{D},f(\mathbf{h}^{\prime},\mathbf{c}^{\prime}\mid u)\right).
\end{split}
\]
% \[
%     \hat{y}^{\prime}_i = LLM_{\mathrm{point}}\left(I,\mathcal{D},f(\mathbf{h}^{\prime},\mathbf{c}^{\prime}\mid u)\right) ~;~
%     \hat{y}^{\prime}_{i_m\succ i_n} = LLM_{\mathrm{pair}}\left(I,\mathcal{D},f(\mathbf{h}^{\prime},\mathbf{c}^{\prime}\mid u)\right) ~;~
%     \hat{y}^{\prime}_{i_1},\hat{y}^{\prime}_{i_2},\cdots,\hat{y}^{\prime}_{i_k} = LLM_{\mathrm{list}}\left(I,\mathcal{D},f(\mathbf{h}^{\prime},\mathbf{c}^{\prime}\mid u)\right).
% \]
Then the output answer will be checked manually, and the valid answers will be utilized for further evaluation, while the invalid answers will be excluded. For more details about the prompts, please refer to the link\footnote{\url{https://github.com/rainym00d/LLM4RS/blob/main/assets/prompts.pdf}}.

\section{Experiments}
In this section, we conduct experiments to evaluate ChatGPT and GPT-3.5s to answer the following research questions:

\textbf{RQ1}: How do these LLMs perform on different ranking capabilities across various recommendation domains?  

\textbf{RQ2}: How do the LLMs-based recommenders compare with traditional collaborative filtering methods?

\textbf{RQ3}: How much cost of these LLMs-based recommenders on different ranking capabilities?

\textbf{RQ4}: How does the number of prompt shots affect the performance of LLMs-based recommenders?

\subsection{Experimental Settings}
\subsubsection{Datasets}
To better probe the different capabilities of ChatGPT and GPT-3.5s on personalized recommendation, we conducted evaluations on datasets from four different domains. 

\textbf{Movie}: We use the widely-adopted MovieLens-1M\footnote{\href{https://grouplens.org/datasets/movielens/1m/}{https://grouplens.org/datasets/movielens/1m/}} dataset  that contains 1M user ratings for movies.

\textbf{Book}: We use the ``Books'' subset of Amazon\footnote{\href{http://jmcauley.ucsd.edu/data/amazon/}{http://jmcauley.ucsd.edu/data/amazon/}\label{amazon}} dataset that contains user ratings for books. 

\textbf{Music}: We use the ``CDs \& Vinyl'' subset of Amazon\textsuperscript{\ref {amazon}} to conduct experiments on the music domain.

\textbf{News}: We use the MIND-small\footnote{\href{https://msnews.github.io/}{https://msnews.github.io/}} dataset as the benchmark for news domain.

Following the common practices \cite{he2017neural, mao2021simplex, xu2022dually}, for the Movie, Book, and Music datasets, we treat ratings above 3 as positive feedbacks (labeled as 1) and otherwise as negative feedbacks (labeled as 0).
For the News dataset, we used the original binary feedback labels. In the experiments, we use the titles of the items as description in the prompt.

\subsubsection{Evaluation Protocols}
After processing, we random sample $500$ records on each dataset for evaluation due to the expensive cost. For all experiments, we follow the existing practice~\cite{sileo2022zero} and pair one positive item with four randomly sampled negative items as the candidate item list. We set the number of shots as 1 for pair-wise and list-wise, and 2 for point-wise. We report top-$K$ Normalized Discounted Cumulative Gain (NDCG@$K$) and Mean Reciprocal Rank (MRR@$K$) with $K=3$.
Furthermore, considering that LLMs may generate some illegal output, that is, results that are not in the candidate set, we introduce the metric ``Compliance Rate'' to compare the compliance rates between different models, which is defined as the proportion of the number of valid results generated to all test samples, i.e., $\frac{\text{Number of Valid Answers}}{\text{Number of Test Samples}}$.

\begin{table*}[]
  \caption{Performance of different LLMs with zero-shot and few-shot examples on Movie dataset. Bold indicates the best result for each row and `\_' indicates the best result for each wise of each LLM. }
  \label{tab:zero_shot}
\resizebox{1\textwidth}{!}{
\begin{tabular}{c|c|c|c|cc|cc|cc}
\toprule
         & & \multirow{2}{*}{random} & \multirow{2}{*}{pop} & \multicolumn{2}{c}{point-wise} & \multicolumn{2}{|c|}{pair-wise} & \multicolumn{2}{c}{list-wise}             \\
\multirow{-2}{*}{Model}            & \multirow{-2}{*}{Metric} & & & zero-shot      & few-shot      & zero-shot      & few-shot     & zero-shot         & few-shot \\
\midrule
         % & NDCG@1    & 0.2000 & 0.2240  & 0.2659    & \underline{0.3110}    & 0.2913    & \textbf{\underline{0.3203}}    & 0.2320    & \underline{0.2600}   \\
         & NDCG@3    & 0.4264 & 0.4761  & 0.5168    & \underline{0.5416}    & 0.5253    & \textbf{\underline{0.5728}}    & 0.4544    & \underline{0.4990}   \\
\multirow{-2}{*}{text-davinci-002}
         & MRR@3   & 0.3667 & 0.4103    & 0.4519    & \underline{0.4824}    & 0.4643    & \textbf{\underline{0.5071}}    & 0.3950    & \underline{0.4363}     \\
\midrule
         % & NDCG@1    & 0.2000 & 0.2240  & \underline{0.2458}    & 0.2259    & \textbf{\underline{0.2903}}    & 0.2843    & 0.2660    & \underline{0.3260}   \\
         & NDCG@3  & 0.4264 & 0.4761     & \underline{0.4674}    & 0.4618    & 0.5249    & \textbf{\underline{0.5441}}    & 0.5062    & \underline{0.5564}   \\
\multirow{-2}{*}{text-davinci-003} 
         & MRR@3   & 0.3667 & 0.4103     & \underline{0.4092}    & 0.3998    & 0.4633    & \textbf{\underline{0.4763}}    & 0.4450    & \underline{0.4950}    \\
\midrule    
         % & NDCG@1    & 0.2000 & 0.2240  & 0.2796    & \textbf{\underline{0.3342}}    & \underline{0.3457}    & 0.3230    &     & 0.3320   \\
         & NDCG@3  & 0.4264 & 0.4761     & 0.5413    & \textbf{\underline{0.5912}}    & \underline{0.5833}    & 0.5827    & \multirow{2}{*}{N/A}        & 0.5785   \\    
\multirow{-2}{*}{gpt-3.5-turbo (ChatGPT)}  
         & MRR@3  & 0.3667 & 0.4103      & 0.4742    & \textbf{\underline{0.5260}}    & \underline{0.5243}    & 0.5162    &         & 0.5167   \\
\bottomrule
\end{tabular}}
\end{table*}

\begin{figure*}[t]
  \subfigure[point-wise]
    {
    \includegraphics[width=0.3\textwidth]{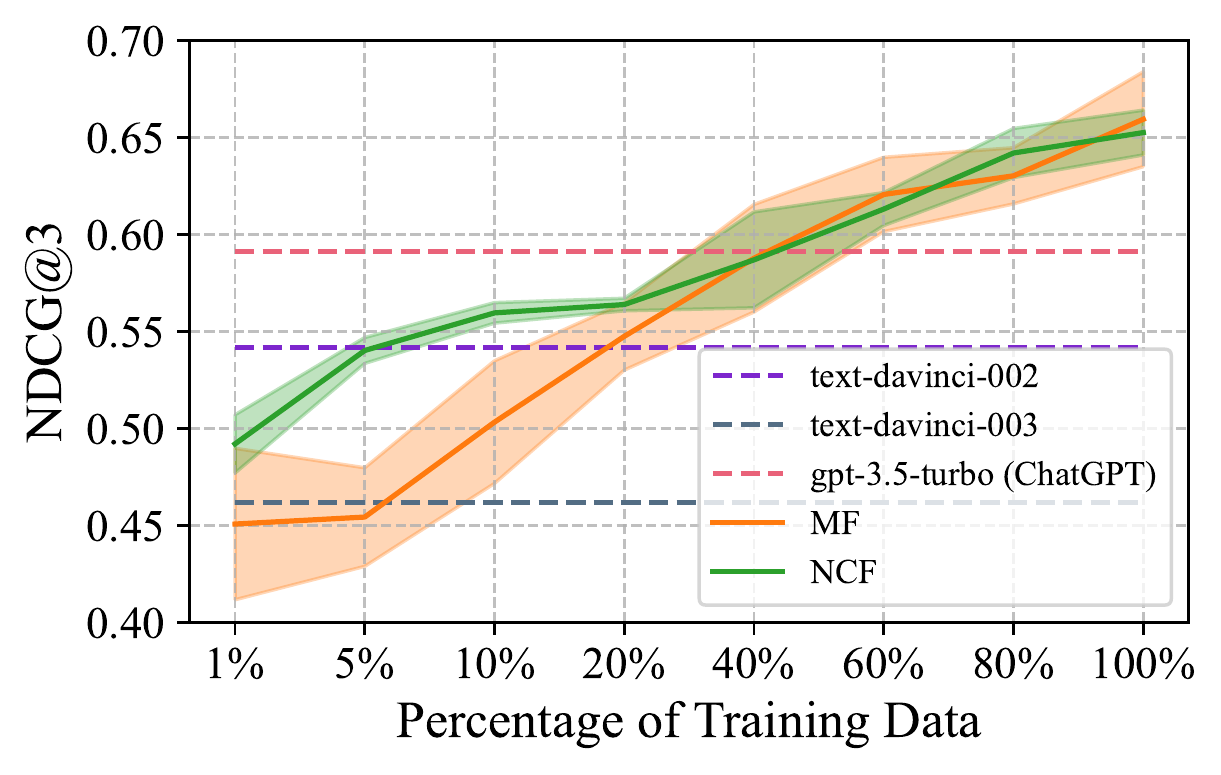}
    \label{fig:mf_point_wise}
   }
   \quad
  \hspace{-0.15in}
  \subfigure[pair-wise]{
    \includegraphics[width=0.3\textwidth]{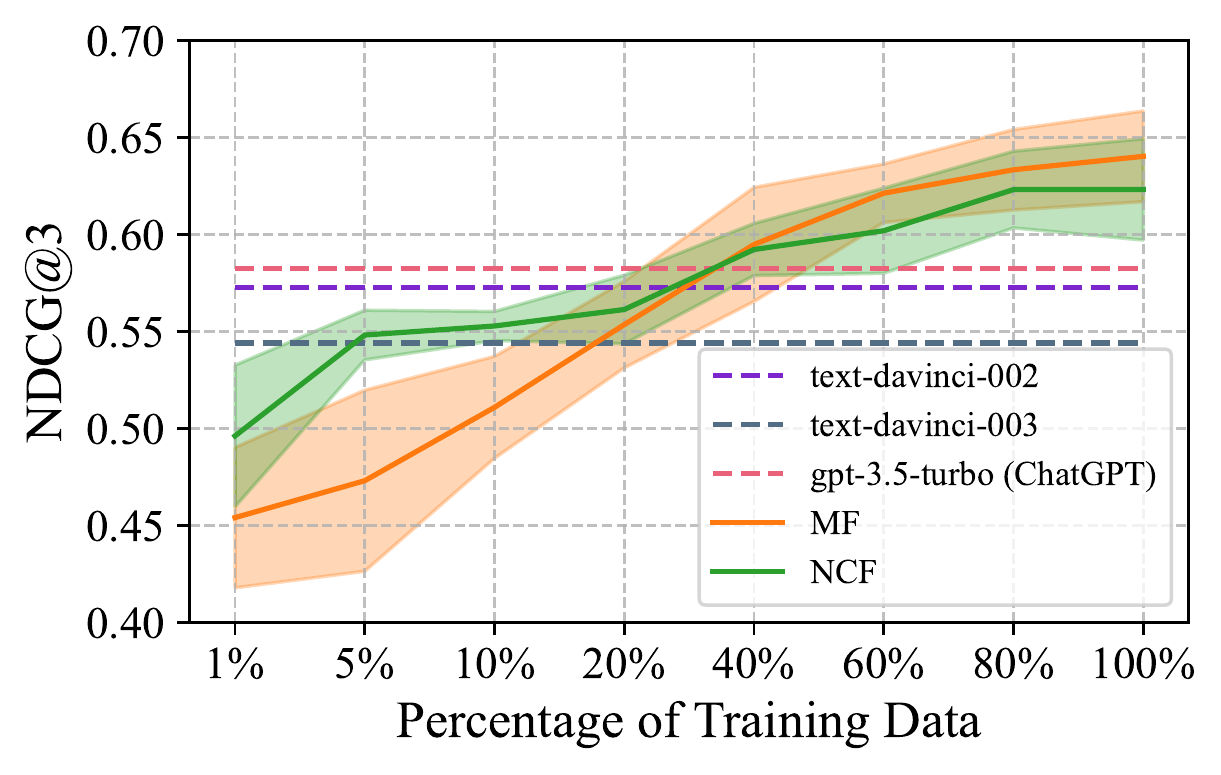}
    \label{fig:mf_pair_wise}
  }
  \quad
  \hspace{-0.15in}
    \subfigure[list-wise]{
    \includegraphics[width=0.3\textwidth]{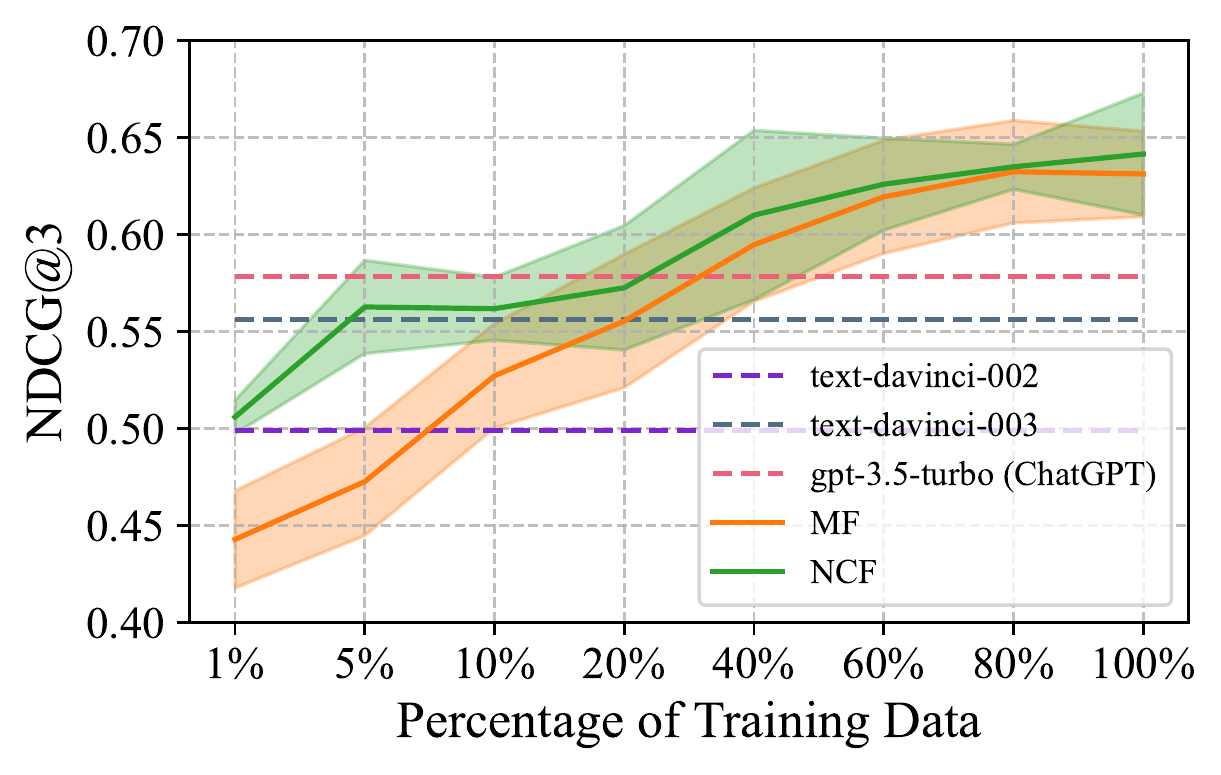}
    \label{fig:mf_list_wise}
  }
    \caption{Comparison with collaborative filtering models in terms of different percentages of training data on Movie dataset. The shaded area indicates the 95\% confidence intervals of $t$-distribution under 5 different experiments with random seeds.}
    \label{fig:analysis_of_mf}

\end{figure*}

\subsection{RQ1: Overall Performance} 
Table \ref{tab:main_results} shows the results of different LLMs on four different domains. We have the following observation and conclusions:

\textbf{ChatGPT and GPT3.5s performed much better than the random recommendation in almost all cases.} 
    Specifically, all three LLMs achieve significant improvements than the random recommendation policy on four domains, e.g.,  average improvements with 24.71\% on the point-wise task in terms of  $NDCG@3$ on the Movie Dataset.
    Additionally, most answers of LLMs are compliant due to the capability of in-context learning. 
    These results reveal that LLMs have the potential to facilitate recommender systems.

\textbf{In comparison to the text-davinci-002 and text-davinci-003, ChatGPT exhibits better performance on almost all evaluation metrics for all three ranking capabilities.} 
    For instance, ChatGPT outperformed the other LLMs in 22 out of 24 comparisons, including two ranking metrics, three ranking capabilities, and four domain datasets. The only two exceptions were for point-wise ranking in the news domain when compared to text-davinci-003. We attribute ChatGPT's strong performance to its exceptional capacity for language understanding and reasoning, which allows it to effectively comprehend item similarity and make informed ranking decisions.

\textbf{ChatGPT performs better with list-wise ranking except in the movie domain. On the other hand, text-davinci-002 and text-davinci-003 perform better with pair-wise ranking in most cases.} To provide a clear comparison, we have summarized the ranking of the three LLMs with different ranking capabilities in Table~\ref{tab:rank}. Note that pair-wise ranking tends to be better than point-wise ranking in most cases (11 out of 12), although it requires more inference cost due to the need for pair-wise comparisons. We will delve deeper into the cost analysis in \textbf{RQ3}.

\textbf{All LLMs-based recommenders outperform the popularity recommendantion policy in recommending movies, books, and music, but they underperform in the news domain.} This phenomenon could be explained by the fact that news recommendation relies more on popularity, while other domains are more personalized. 
    The speed of news delivery is another possible factor.
    Due to the time-sensitive and rapidly changing nature of news recommendation, there is often insufficient interaction data available for each news in the LLMs training corpus.
    Conversely, in the other three domains, the item descriptions and interaction data are more abundant, making LLMs works better on them.
    Overall, this observation suggests that while off-the-shelf LLMs-based recommenders can be effective in many domains, they may not be suitable for some domain and may require further exploration.

We also conduct experiments using zero-shot prompts (i.e., without examples). However, with the original zero-shot prompt, we find more than 50\% of cases were invalid and challenging to evaluate. To address this, we utilize logit\_bias\footnote{\url{https://platform.openai.com/docs/api-reference/completions/create\#completions/create-logit_bias}} to control the output tokens. Due to the page limitation, we provide the detailed results in the link\footnote{\url{https://github.com/rainym00d/LLM4RS/blob/main/assets/Supplementary_Material.pdf}}. Overall, the results highlight the potential of LLMs as recommendation systems, as they outperform random and popularity-based policies in the zero-shot setting. Furthermore, as expected, LLMs under few-shot settings outperform those under zero-shot settings in most cases, demonstrating the effectiveness of few-shot prompts in-context learning.

\begin{figure*}[tp]
    \centering    
    \subfigure[Movie]
    {
        \includegraphics[width=0.23\linewidth]{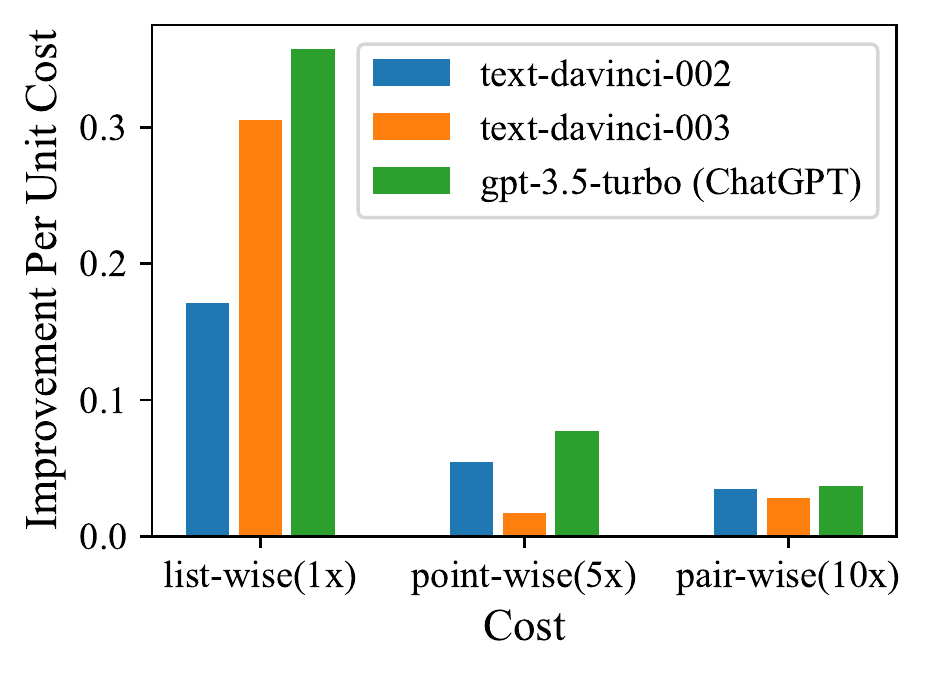}
    }
    \subfigure[Book]
    {
        \includegraphics[width=0.23\linewidth]{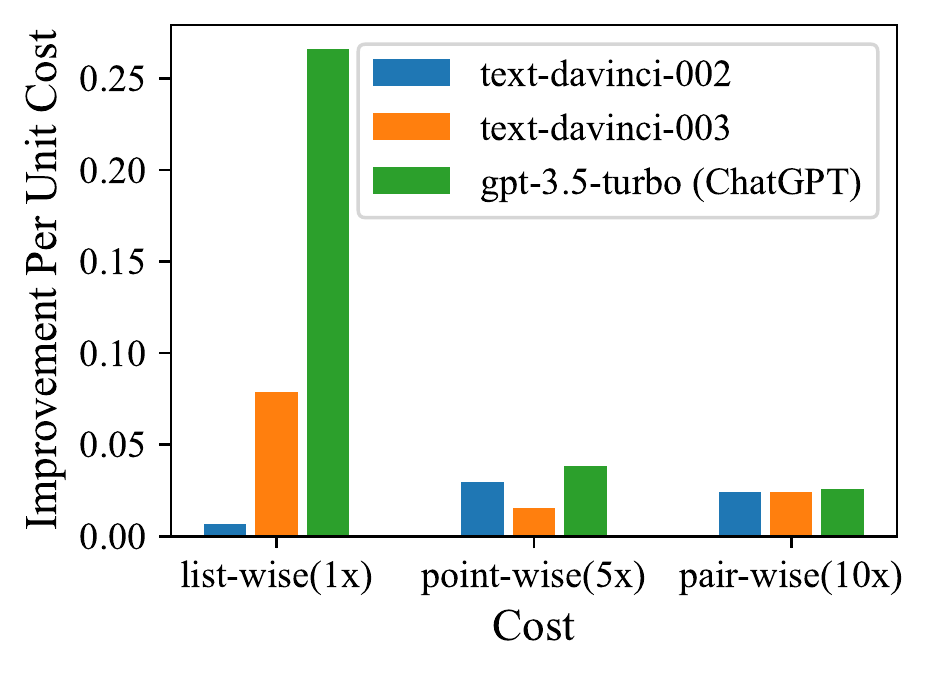}
    }
    \subfigure[Music]
    {
        \includegraphics[width=0.23\linewidth]{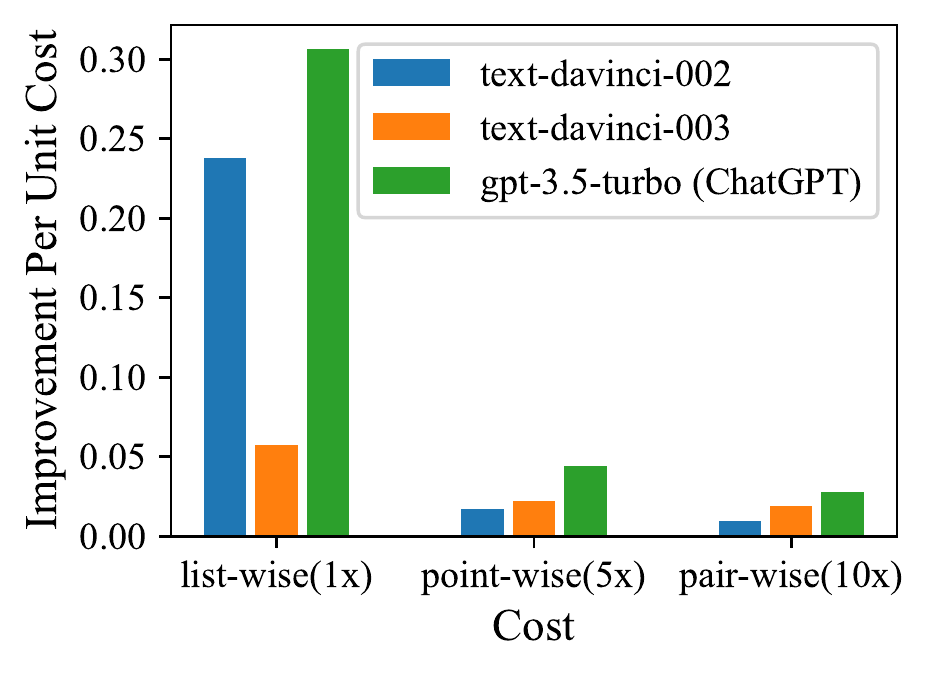}
    }
    \subfigure[News]
    {
        \includegraphics[width=0.23\linewidth]{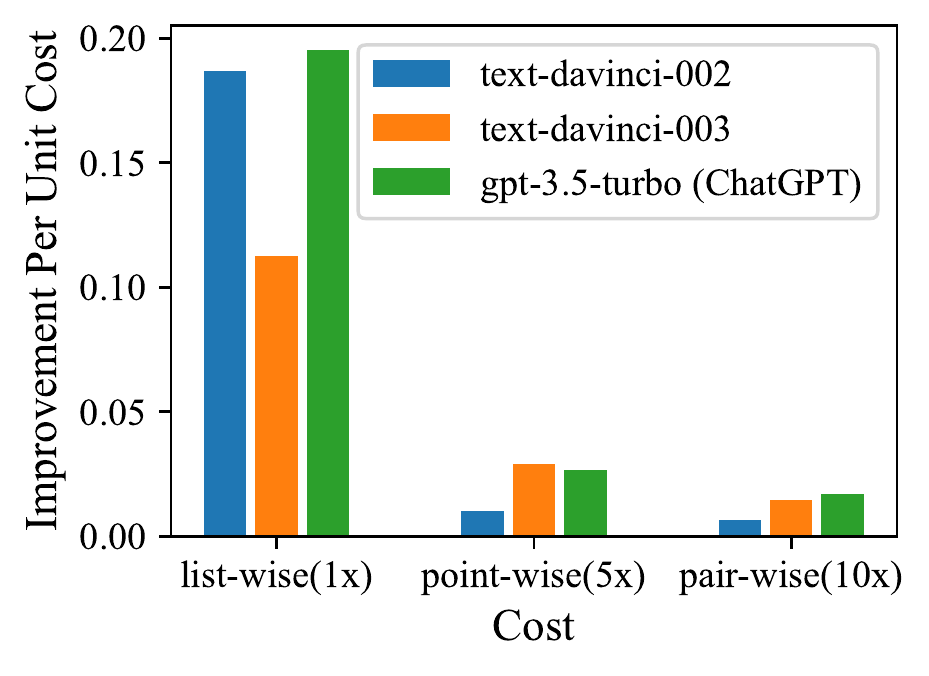}
    }
    \caption{Improvement of $NDCG@3$ per unit cost and five shots examples on four datasets. `1x 5x 10x' denote the cost of list-wise, point-wise, and pair-wise, respectively.}
    \label{fig:analysis_COST}
\end{figure*}

\begin{figure*}[t]
  \subfigure[text-davinci-002]
    {
    \includegraphics[width=0.3\textwidth]{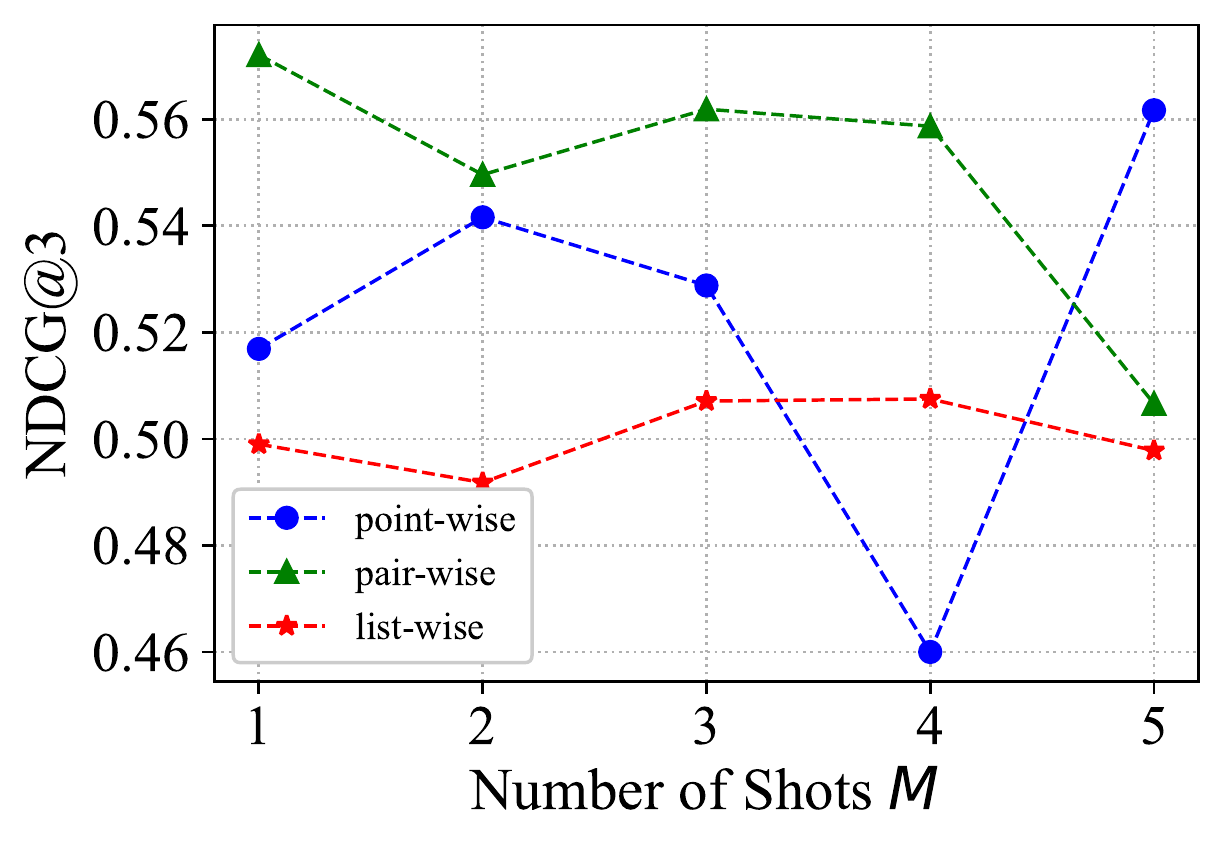}
    \label{fig:002_shots}
   }
  % \hspace{-0.05in}
  \subfigure[text-davinci-003]{
    \includegraphics[width=0.3\textwidth]{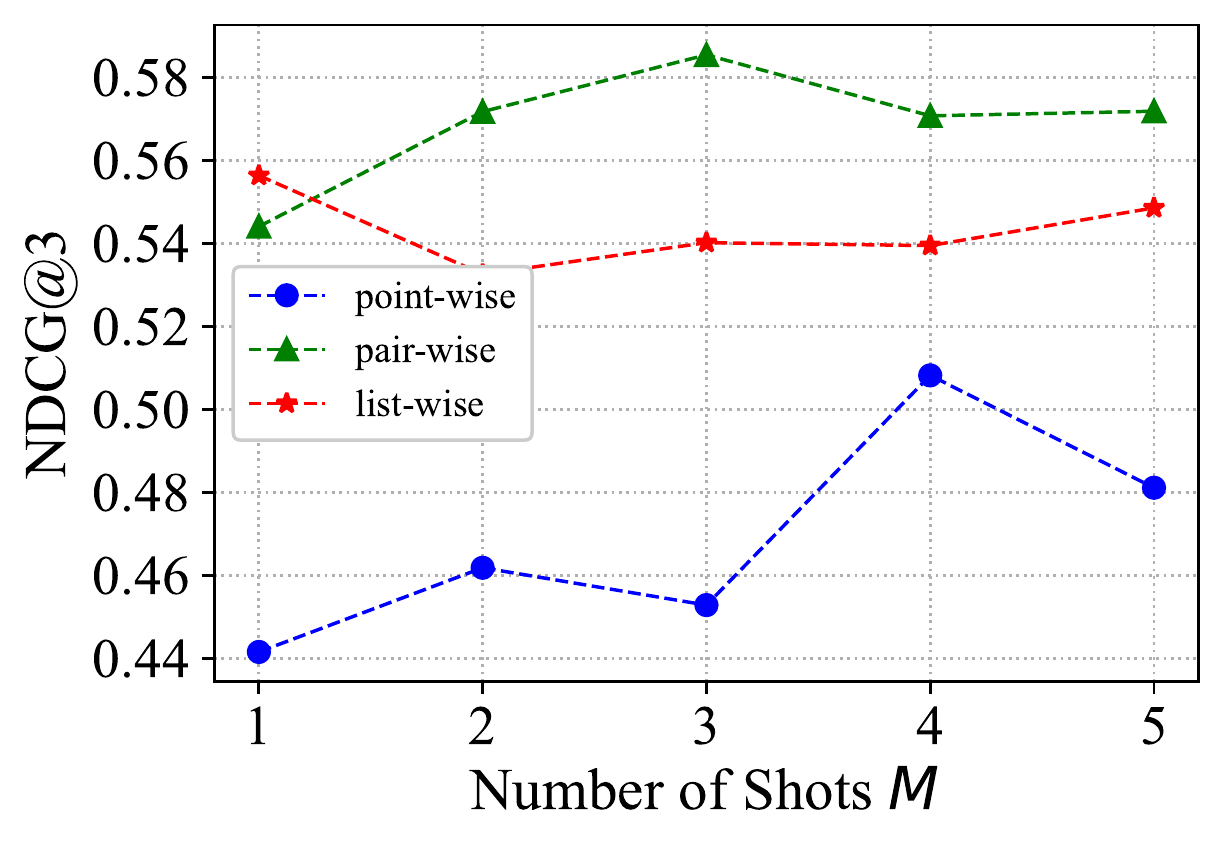}
    \label{fig:003_shots}
  }
  % \hspace{-0.05in}
    \subfigure[gpt-3.5-turbo (ChatGPT)]{
    \includegraphics[width=0.3\textwidth]{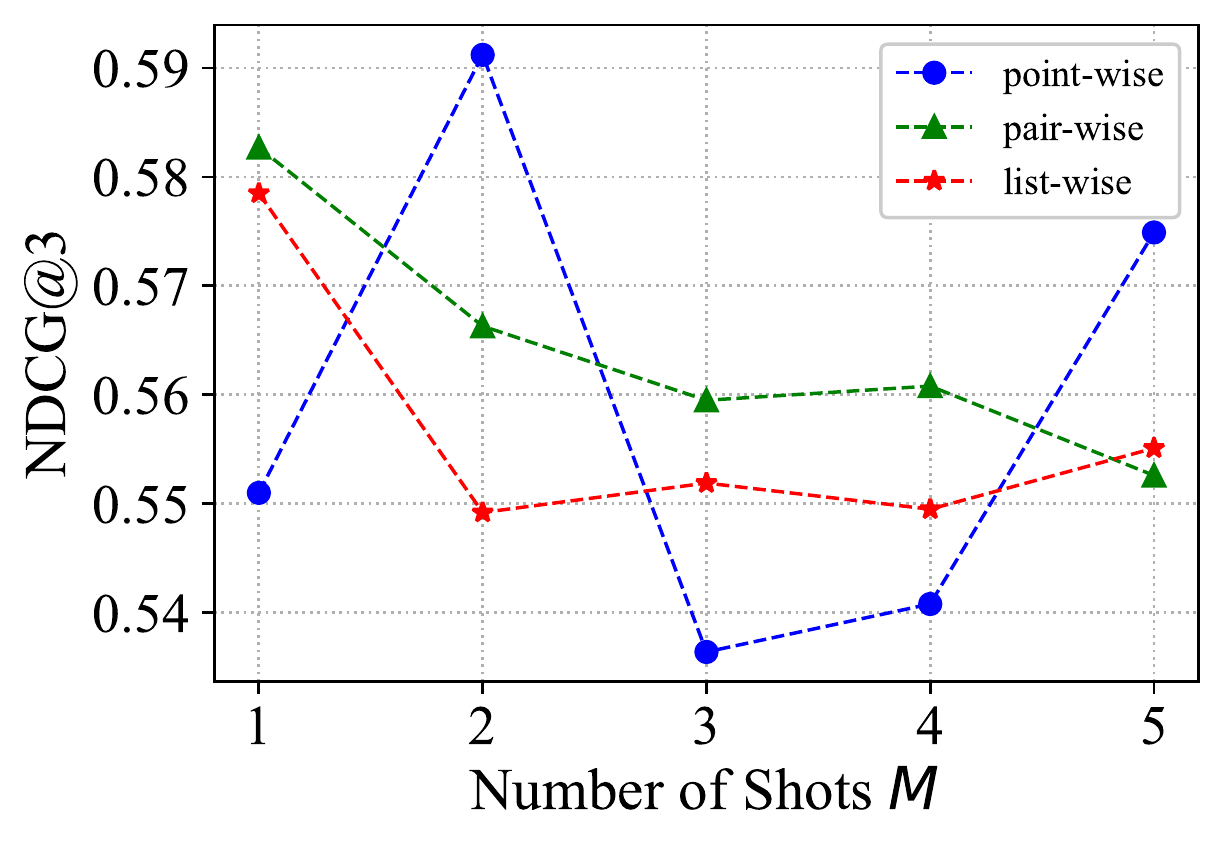}
    \label{fig:chatgpt_shots}
  }
    % \vspace{-1em}
    \caption{Impact of the number of shots prompts in LLMs on Movie dataset.}
    \label{fig:analysis_of_shots}
\end{figure*}

\subsection{RQ2: Comparison with Collaborative Filtering Models}

Given that the LLMs used in the previous experiments were not trained on recommendation data, we investigate the amount of training data required for traditional recommendation models to achieve performance comparable to or better than LLMs.
Specifically, we chose the most representative traditional recommendation models, Matrix Factorization (MF)~\cite{koren2009matrix} as well as Neural Collaborative Filtering (NCF)~\cite{he2017neural}, and evaluated their performance after training on varying proportions of data. For a fair comparison, we carefully tune the parameters of MF and NCF. We then compared their performance to that of LLMs. All experiments are conducted $5$ times on the Movie dataset, and the averaged results and their 95\% confidence intervals of $NDCG@3$ were illustrated in Figure~\ref{fig:analysis_of_mf}. 
As expected, the performance of MF and NCF improves with increasing amounts of training data.
Also, we can observe that off-the-shelf LLM-based models outperform MF and NCF when there are only a few training data available, i.e., less than $40\%$ for ChatGPT with all three ranking capabilities.
Note that LLM-based recommendation models do not require training data but rather a few samples in the prompt to help understand the recommendation task.  
Therefore, we conclude that LLM-based recommendation models can be applied  in practice to mitigate the cold start problem.

\subsection{RQ3: Performance Scaling by Cost}
% (performance - random / random)/ cost 画的图表示的是单位成本的improvement
Although the LLMs have better performance on pair-wise or list-wise ranking as presented in Table~\ref{tab:main_results}, we need to consider the costs associated with  these performance improvements. Specifically, we calculate the improvement per unit cost for each LLM: $\frac{\frac{V_{LLM} - V_{random}}{V_{random}}}{cost_{LLM}}$, where $V_{LLM}$ denotes the metric value of the LLM, $V_{random}$ denotes the metric value of random recommendation, $cost_{LLM}$ denotes the cost of ranking one user's candidate item list.
Referring to Figure~\ref{fig:model} (left), let us define the $cost_{LLM}$. 
For list-wise ranking, only one prompt input is needed to obtain LLM's ranking for all candidate items. For point-wise ranking, N prompt inputs are required to obtain LLM's ranking for all candidate items (where N is the number of candidate items). For pair-wise ranking, $\frac{N(N-1)}{2}$ prompt inputs are required to obtain all ranking results.
In our experimental settings, $N$ is set to 5. The costs of point-wise, pair-wise, and list-wise ranking are denoted as \textbf{5x}, \textbf{10x} and \textbf{1x}, respectively. 
Figure~\ref{fig:analysis_COST} demonstrates the improvement per unit cost of each LLMs. It can be found that almost all three LLMs has the best improvement per unit cost in list-wise ranking, except text-davinci-002 on the Book dataset.
Moreover, point-wise ranking and pair-wise ranking have similar improvement per unit cost. 
Although pair-wise ranking may achieve better performance than point-wise ranking in absolute metrics, the requirement to run multiple prompts for pair-wise ranking results in additional cost.
Overall, we recommend to conduct list-wise ranking for recommendation tasks in practice, due to its decent performance and lower cost.

\subsection{RQ4: Performance Under Different Shots Examples}
Previous studies in NLP have emphasized  that the number of examples $M$ is important for in-context learning. To assess  the impact of $M$ in LLMs for recommendation, we conducted experiments  on Movie dataset by varying $M$ from $1$ to $5$. Figure~\ref{fig:analysis_of_shots} illustrates the performances of different $M$ in terms of $NDCG@3$ of ChatGPT and GPT3.5s.  Surprisingly, we observe that the best results did not always correspond to the maximum number of examples. One possible explanation is that while more example shots can provide more context and information for LLMs to understand the recommendation task, they may also introduce more noise, causing LLMs to learn unhelpful patterns. Therefore, the optimal number of prompt shots may depend on the specific LLM, task, and dataset.

\begin{table*}[t]
  \caption{Case Study of Exceptions. The \textcolor{Answer}{green} is the answer of ChatGPT.}
  \label{tab:case_study}
    \resizebox{0.9\textwidth}{!}{
    \begin{tabular}{p{6 cm}|p{9 cm}}
    \toprule
\multicolumn{1}{c|}{Case 1}     & \multicolumn{1}{c}{Case 2}   \\
You are a movie recommender system now.    & You are a book recommender system now.   \\
$\{\{Examples\}\}$         & $\{\{Examples\}\}$       \\
Input: Here is the watching history of a user: Aliens, E.T. the Extra-Terrestrial, Contact, The Matrix, X-Men. Based on this history, would this user prefer The Fox and the Hound or Steamboat Willie? Answer Choices: (A) The Fox and the Hound (B) Steamboat Willie & Input: Here is the reading history of a user: The Cellist of Sarajevo, After I'm Gone: A Novel, The Reason I Jump: The Inner Voice of a Thirteen-Year-Old Boy with Autism, The Serpent of Venice: A Novel, We Are All Completely Beside Ourselves: A Novel. Based on this history, would this user prefer The Secret Life of Bees or The Help? Answer Choices: (A) The Secret Life of Bees (B) The Help \\
Output: The answer index is \textcolor{Answer}{N/A as neither option is relevant to the user's watching history.}  & Output: The answer index is \textcolor{Answer}{N/A. It is difficult to determine the user's preference based on this reading history as neither book is similar in genre or theme to the books they have read.} \\
    \bottomrule
    \end{tabular}}
\end{table*}

\subsection{Case Study of Exceptions}

It is worth noting that the LLM may generate some invalid answers even under few-shot in-context learning, leading to a compliance rate may be less than 100\%, as shown in Table~\ref{tab:main_results}. For instance, Table~\ref{tab:case_study} highlights two exceptional cases of answers from ChatGPT with pair-wise ranking, where both cases lack a correct answer because they are the pair of two negative samples. Surprisingly, ChatGPT does not simply respond with `A' or `B' as seen in the in-context learning examples. Instead, it recognizes that these two items are unrelated and not similar to the user history interactions.
For example, in case 1, the user watching histories are all science fiction movies but the answer choices are all cartoons. These responses demonstrate that ChatGPT can understand how to recommend based on the user interaction histories and what is the similarity between items. However, limited by our existing evaluation methods, these answers are considered non-compliant. 
Therefore, we suggest exploring additional perspectives for evaluating LLMs as recommenders beyond learning to rank, as LLMs have the potential to play a larger role in explainable recommendation.

\section{Conclusion}
In this paper, we conduct a preliminary evaluation for probing off-the-shelf LLMs for recommendation from the point-wise, pair-wise, and list-wise perspectives. Specifically, we reformulate the above ranking capabilities into corresponding domain-aware prompts and evaluate the performance of ChatGPT in each ranking capability on different domains. The results on four datasets indicate the superiority of ChatGPT in recommendations among all three ranking capabilities. We also observe that LLMs excel at list-wise and pair-wise ranking, but are not proficient in point-wise ranking in most cases. Furthermore, ChatGPT
shows the potential for mitigating the cold start problem and explainable recommendation.

\begin{acks}
This work was funded by the National Key R\&D Program of China (2019YFE0198200), Beijing Outstanding Young Scientist Program NO. BJJWZYJH012019100020098, Intelligent Social Governance Interdisciplinary Platform, Major Innovation \& Planning Interdisciplinary Platform for the ``Double-First Class'' Initiative, Renmin University of China. Supported by fund for building world-class universities (disciplines) of Renmin University of China.
\end{acks}

\balance
\bibliographystyle{ACM-Reference-Format}
%\bibliography{sample-base}
%\bibliographystyle{plain}
\bibliography{ref}

\end{document}